\documentclass[twocolumn]{aastex63}
\usepackage{amsmath,graphicx,epsfig}
\usepackage{lipsum}
\usepackage{bigstrut,bigdelim,multirow}

\newcommand{\be}{\begin{eqnarray}}
\newcommand{\ee}{\end{eqnarray}}

\begin{document}
\title{An accreting stellar binary model for active periodic fast radio bursts}
%\title{A X-ray Binary Model for Repeating Fast Radio Bursts}
\author{Can-Min Deng}
%\email{Corresponding author: dengcm@gxu.edu.cn}
\affil{GXU-NAOC Center for Astrophysics and Space Sciences, Department of Physics, Guangxi University, Nanning 530004, China; dengcm@gxu.edu.cn}
\affil{CAS Key Laboratory for Research in Galaxies and Cosmology, Department of Astronomy, University of Science and Technology of China, Hefei 230026, Anhui, China}
\author{Shu-Qing Zhong}
\affil{CAS Key Laboratory for Research in Galaxies and Cosmology, Department of Astronomy, University of Science and Technology of China, Hefei 230026, Anhui, China}
%\affil{School of Astronomy and Space Science, Nanjing University, Nanjing 210093, China}
%\affil{Key laboratory of Modern Astronomy and Astrophysics (Nanjing University), Ministry of Education, Nanjing 210093, China}
\author{Zi-Gao Dai}
\affil{CAS Key Laboratory for Research in Galaxies and Cosmology, Department of Astronomy, University of Science and Technology of China, Hefei 230026, Anhui, China}

\begin{abstract}
In this work, we propose an accreting stellar binary model for understanding the active periodic fast radio bursts (FRBs). The system consists of a stellar compact object (CO) and a donor star (DS) companion in an eccentric orbit, where the DS fills its own Roche lobe near the periastron. The CO accretes the material from the DS and then drive relativistic magnetic blobs.  The interaction between the magnetic blobs and the stellar wind of the DS produces a pair of shocks. We find that both of the reverse shock and the forward shock are likely to produce FRBs via synchrotron maser mechanism. We show that this system can in principle sufficiently produce highly active FRBs with a long lifetime, and also can naturally explain the periodicity and the duty cycle of the activity as appeared in FRBs 180916 and 121102. The radio nebula excited by the long-term injection of magnetic blobs into the surrounding environment may account for the associated persistent radio source. In addiction, we discuss the possible multi-wavelength counterparts of FRB 180916  in the context of this model.  Finally, we encourage the search for FRBs in the ultraluminous X-ray sources.

\end{abstract}


\section{Introduction}
\label{sec:introduction}
Fast radio bursts (FRBs) are intense radio transients  with extremely short duration,  and their physical origin is a mystery \citep{2019ARA&A..57..417C,2019A&ARv..27....4P,2020Natur.587...45Z,2021SCPMA..6449501X}.  Observationally, some FRBs showed repeat bursts, but most bursts did not \citep{2016Natur.531..202S,2020ApJ...891L...6F,2020MNRAS.495.2416J}. It is also a mystery whether all of FRBs have the same origin \citep{2018ApJ...854L..12P,2018NatAs...2..839C,2019MNRAS.484.5500C}.

\cite{2020Natur.582..351C} found a 16.35-day periodicity in FRB 180916 with a 4.0-day phase window, implying a duty cycle $\rm{D} \simeq 0.24$ of activity.  In addiction, \cite{2020MNRAS.495.3551R} reported a possible 157-day periodicity in the FRB 121102 with a duty cycle $\rm{D} \simeq 0.56$.
These are very important clues for studying the physical origin of the FRBs.
For such a periodicity, it may be explained, either by the orbital period of the binary sysytem  \citep{2020ApJ...890L..24Z,2020ApJ...893L..39L,2020ApJ...893L..26I,2020MNRAS.497.1543G,2020A&A...644A.145M,2020ApJ...895L...1D}, or precession of the emitter  \citep{2020ApJ...893L..31Y,2020ApJ...895L..30L,2020ApJ...892L..15Z,2020RAA....20..142T,2020MNRAS.494L..64K}.

Recently, \cite{2020Natur.587...54C} and \cite{2020Natur.587...59B} reported a single  FRB (FRB 200428) with two pulses in association with an active Galactic magnetar SGR 1935+2154, which is located at a distance $\sim 9$ kpc \citep{2020ApJ...898L...5Z}.
Moreover, it is found that an X-ray burst is almost simultaneous with the FRB \citep{2020arXiv200511071L,2020ApJ...898L..29M,2020arXiv200511178R,2020arXiv200512164T}.
But strangely,  no more FRBs were observed during the active phase of this magnetar \citep{2020Natur.587...63L}.
In any case, this discovery shows that at least a part of the FRBs are produced by the magnetars \citep{2014MNRAS.442L...9L,2017ApJ...843L..26B,2019MNRAS.485.4091M}. However, an important lesson learned from FRB 200428 is that extragalactic analogues of Galactic magnetars could explain some of the FRB population, but much more active sources are still required to be invoked to explain the highly active periodic repeaters like FRBs 121102 and 180916 \citep{2020Natur.587...54C}. 
If the active repeaters are also powered by magnetars,
they must be produced by a type of rare and active magnetars not seen in the Milky Way \citep{2020ApJ...898L..29M,2020MNRAS.498.1397L}.

In addition, the observations on the environment of FRB 180916 found that it is 250 pc away from the nearest young stellar clump and suggested that its age may be $\sim$ Myr, which in line with the hypothesis that the high-mass X-ray binaries or gamma-ray binaries are the progenitors rather than the young magnetars \citep{2020arXiv201103257T}.  Moreover, \cite{2020arXiv201208348P} rule out the scenario in which companion winds cause FRB periodicity by using simultaneous Apertif and LOFAR data.
Motivated by those observation results mentioned above, we propose an alternative model for understanding the highly active repeating FRBs with periodicity in this work.
The model consists of a stellar compact object (CO) and a donor star (DS) with its filled Roche lobe, in which the CO can be a neutron star (NS) or a black hole (BH). We will show that this model can explain the actively repeating FRBs themselves and their periodicity behaviors, as the cases FRBs 180916 and 121102.

{\cite{2021arXiv210206138S} also link the periodic FRBs to the accreting binaries. In \cite{2021arXiv210206138S}, the FRBs are produced by  the interaction of the intermittent jets (high luminosity)with the quiescent jet (low luminosity).
However, in this work, we discuss the process by which magnetic blobs interact with the  donor star's wind to produce FRBs. This is the key difference between this work and theirs. 
}

\section{The model}
\begin{figure*}
	\centerline{\includegraphics[angle=0,width=1\hsize]{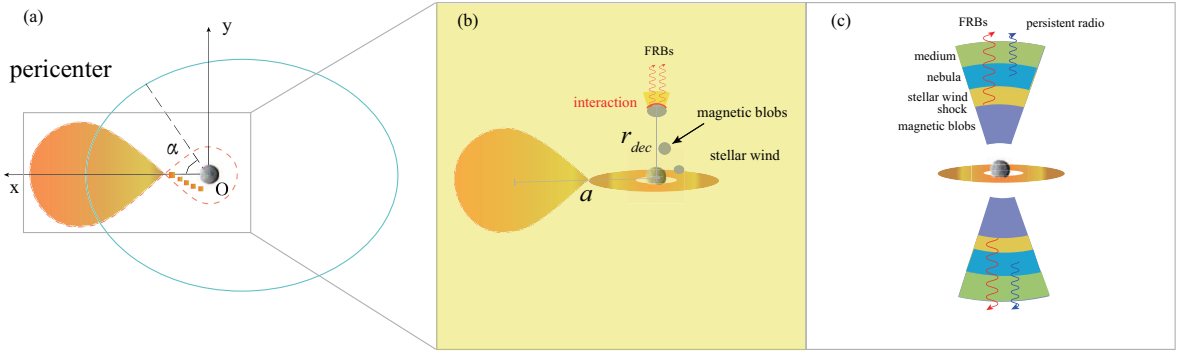}}
	\caption{Schematic illustration of the model in this work: (a) the DS fills its Roche lobe when orbiting near the periastron, and mass transfer to the CO occurs; (b) The transfer of mass leads to the formation of accretion disc around the CO. Magnetic blobs are ejected from the accretion disc due to magnetic instability. The magnetic blobs may accelerate to the Lorentz factor $\Gamma \sim \sigma_0$ \citep{2012ApJ...757...56Y}, where $\sigma_0$ is the initial magnetization parameter of the blobs.  Then the blobs  interact with the stellar wind, which is driven by the DS and immerses itself around the binary system, to induce shocks powering FRBs by the synchrotron maser mechanism; (c) The interactions of the magnetic blobs and the stellar wind produce FRBs. Long term energy injection into the  surrounding medium procduce a nebulae which powers the persistent raido source associated with the FRB. 
	}
	\label{model}
	~\\
\end{figure*}

In this section, we propose a binary model for active repeating FRBs.
The binary system consists of a stellar CO and a DS. The CO may be a BH or a NS.
As illustrated in Figure \ref{model}, when the DS fills its Roche lobe, significant mass transfer will occur from the DS to the CO.
Then an accretion disc will form around the CO.
As we know, jets are widely present in the accretion systems. However, in addition to the usual continuous jets, the accretion process may also produce episodic jets.
The energetic, collimating episodic jets had been observed in active galactic nuclei, stellar  binaries, and protostars. For instance, in the X-ray binaries, episodic jets are usually observed during their X-ray outbursts and intense
radio fares \citep{2004MNRAS.355.1105F,2015MNRAS.451.1740Z}. 
Practical models and numerical simulation suggested that the episodic jets may be driven by magnetic instability in the accretion disc \citep{2009MNRAS.395.2183Y,2012ApJ...757...56Y,2020MNRAS.499.1561Z}. Because of shear and turbulent motion of the accretion flow, a flux rope system is expected to form near the disc.
The energy is accumulated and stored in the system until a threshold is reached,  then the system loses its equilibrium and  the energy will be released in a catastrophic way, i.e. ejecting episodic magnetic blobs. 
By assuming the accretion flow is advection dominated, the available isotropic free magnetic energy of one blob is \citep{2012ApJ...757...56Y}
\begin{equation}
E_{\rm{f}} \sim 10^{42}~ f_{\rm{b,-3}}^{-1} \alpha^{-1}_{-2} \beta_{-1}\left(\frac{{{\dot{M}_{a}}}}{10^{-5} M_{\odot} / \rm{yr}}\right) \left(\frac{{{M}}}{ 3M_{\odot} }\right)~\hat{r}_{\rm{2}}^{1 / 2}~ \rm{erg}.
\label{eq:Ef}   
\end{equation}	
Here we consider the magnetic blobs is ejected in a collimated angle, 
{and  $f_{\rm{b}}=(1-\cos \theta)/2$ is the beaming factor, $\theta$ is the jet opening 
angle\footnote{
{In the context of this model, the jets are ultra-relativistic, as shown in Eq.(\ref{eq:gamma}). And the ultra-relativistic jets are usually collimated. Considering the jets with an opening angle $\theta$ equivalent to the reciprocal of the Lorentz factor $\Gamma$ i.e. $\theta\sim 1/ \Gamma$, we have $f_b \sim 1/ \Gamma^{2} \sim O(10^{-4}-10^{-3})$ according to Eq.(\ref{eq:gamma}).}
}.  Here, we consider a beaming factor $f_{\rm{b}} \sim 10^{-3}$.} 
$\alpha$ is the viscous parameter, $\beta$ is the ratio of the magnetic pressure over the total pressure  in the accretion disc. We adopt $\alpha=0.01$ and $\beta=0.1$ as typical values. $\dot{{M_{a}}}$ is the mass accretion rate, $M$ is the mass of the CO, $\hat{r}$ is the radius, where the magnetic blobs is formed, in units of $2GM/c^2$. 
Here and hereafter, we employ the short-hand notation $q_{x}=q / 10^{x}$ in cgs units.
%is equivalent to an opening angle of 3.6 degrees.

On the other hand, the binary system is surrounded by the stellar wind from the DS.
Therefore, we suggest that interaction between the magnetic blobs and the stellar wind could lead to shock formation, and then this shock process would produced powerful coherent radiation (FRBs) through the synchrotron maser mechanism analogy to the flaring magnetar models \citep{2014MNRAS.442L...9L,2017ApJ...843L..26B,2019MNRAS.485.4091M}. 
Taking FRB 180916 as a template, the typical energy of the bursts is $\sim 10^{37}-10^{38}$ erg \citep{2020Natur.587...54C}. 
{In fact, it turns out that the typical energy of FRB 121102 and of FRB 180916 are similar,  found  by recent observations \citep{2020arXiv201208348P,2021arXiv210708205L}.
Also one will see from Eq.(\ref{eq:Evm}), $E_{\rm{f}}=10^{42}$ erg is reasonable to power FRB 180916 and 121102.   According to equation (\ref{eq:Ef}), a mass accretion rate $\sim 10^{-5} M_{\odot}/\rm{yr}$ is required.
}
Note that this high accretion rate $\dot{{M_a}} \sim 10^{-5} M_{\odot}/\rm{yr}$ can occur in high mass  Roche lobe filling binaries, such as SS433 which is accreting at a rate of $\sim 10^{-4} M_{\odot}/\rm{yr}$ \citep{2004ASPRv..12....1F}. Moreover, \cite{2015ApJ...810...20W} showed that the Roche lobe overflow rate can be up to $10^{-3} M_{\odot} / \rm{yr}$ in if the DS is evolving in the Hertzsprung gap.

The  wind emerged from the DS is filled around the CO, and its density distribution $n_{\rm{w}}$ can be estimated as
\begin{equation}
n_{\rm{w}} = \frac{\dot{M}_{\rm{w}}}{4 \pi m_{\rm{p}} R^{2} v_{\rm{w}}}, ~~R \simeq \left(r^{2}+a^{2}\right)^{1 / 2}
\end{equation}
where $\dot{M}_{\rm{w}}$ is the wind mass loss rate of the DS, $v_{\rm{w}}$ is the speed of the wind. $a$ a is the semimajor axis of the binary, $r$ is the distance from the CO.
For $r\ll a$, we have
\begin{equation}
n_{\rm{w}} \simeq 1.0 \times 10^3 ~ \dot{m}_{\rm{w,-11}}  \beta_{\rm{w,-2}}^{-1} a_{13}^{-2} ~\rm{cm}^{-3}
\label{eq:nw}
\end{equation}	
where $\dot{m}_{\rm{w}}=\dot{M}_{\rm{w}}/( M_{\odot} / \rm{yr})$, $\beta_{\rm{w}}=v_{\rm{w}}/c$, $c$  is the speed of light. We adopt  $\beta_{\rm{w}}=0.01$ as typical values for massive stars, and adopt $\dot{m}_{\rm{w}}=10^{-11}$ for self-consistent of the working model.  $a=10^{13}$ cm is motivated by the analysis for FRB 180916 in section 3.
%And it might be true, noting that \cite{1996ApJ...459..717K} found a strong constraint on the stellar wind of a B-star $<10^{-11} \mathrm{M}_{\odot}/\mathrm{yr}$.

Therefore, according to \cite{1995ApJ...455L.143S}, the Lorentz factor of the blastwave during the early reverse shock crossing phase is $\Gamma=(n_{\mathrm{ej}}\Gamma_{\mathrm{ej}}^{2} /4n_{\mathrm{w}})^{1 / 4}$, where $n_{\mathrm{ej}} \simeq E_{\rm{f}}/(4\pi r^{2}m_{\rm{p}}c^{3} \delta t \Gamma_{\mathrm{ej}}^{2}) $ is the co-moving density in the  ejecta at a radius $r$, $\Gamma_{\mathrm{ej}}$ is the initial Lorentz factor of the  ejecta, $\delta t$ is the duration of the central energy activity. Then we have 
\begin{equation}
\Gamma\left(r < r_{\mathrm{dec}}\right)=\left(\frac{E_{\rm{f}} \beta_{\mathrm{w}}}{ 4\dot{M}_{\rm{w}} c^{2} \delta t}\right)^{1 / 4}\left(\frac{r}{a}\right)^{-1 / 2},
\end{equation}
where 
\begin{equation}
r_{\mathrm{dec}}  \simeq 1.1 \times 10^{12} E_{\rm{f,42}}^{1 / 4} \dot{m}_{\mathrm{w,-11}}^{-1 / 4} \beta_{\mathrm{w,-2}}^{1/4}  \delta t_{-3}^{1 / 4}a_{13}^{1/2}~ \mathrm{cm}
\label{eq:rdec}
\end{equation}
is the deceleration radius. One sees that it satisfies $r_{\mathrm{dec}} \ll a$. Therefore, the Lorentz factor at the deceleration radius is
\begin{equation}
\Gamma \left(r_{\mathrm{dec}}\right) \simeq 136~ E_{\rm{f,42}}^{1 / 8} \dot{m}_{\mathrm{w,-11}}^{-1 / 8} \beta_{\mathrm{w,-2}}^{1 / 8} \delta t_{-3}^{-3 / 8} a_{13}^{1/4}~.
\label{eq:gamma}
\end{equation}
{On the other hand, we get  the co-moving density in the  ejecta at $r_{\mathrm{dec}}$}
\begin{equation}
n_{\rm{e j}} \simeq 1.4 \times 10^{8}~ \Gamma_{\rm{e j}, 2}^{-2} E_{\rm{f,42}}^{1 / 2} \dot{m}_{\mathrm{w,-11}}^{1 / 2} \beta_{\mathrm{w,-2}}^{-1 / 2} \delta t_{-3}^{-3 / 2} a_{13}^{-1} \mathrm{~cm}^{-3} .
\label{eq:nej1}
\end{equation}
{{Combining Eq.(\ref{eq:nw}) with Eq.(\ref{eq:nej1})}, one gets that the reverse shock will become relativistic,  which is the case considered in this work, if }
\begin{equation}
\Gamma_{\rm{e j}} \gtrsim 194~  E_{\rm{f,42}}^{1 / 8} \dot{m}_{\mathrm{w,-11}}^{-1/ 8} \beta_{\mathrm{w,-2}}^{1 / 8} \delta t_{-3}^{-3 / 8} a_{13}^{1/4}  ,
\end{equation}
{according to the condition $\Gamma_{\rm{e j}}^{2} > f \equiv n_{\rm{e j}}/n_{\rm{w}}$ \citep{1995ApJ...455L.143S}.  
In the context of this model, the reverse shock is ultrarelativistic only for $\Gamma_{\rm{e j}} \gg 10^{2}$, and is mildly relativistic for $\Gamma_{\rm{e j}} \sim 200$ which is the easier case to achieve.  
{Can accreting stellar-mass compact objects produce ejecta with $\Gamma_{\rm{e j}} >100$? As argued in \cite{2021arXiv210206138S}, it cannot be excluded that highly super-Eddington systems are capable for brief periods of generating outflows with such a large $\Gamma_{\rm{e j}}$. Especially for ejecta with a large initial magnetization $\sigma_{0} \gg 1$, which is the case in this work, it can be accelerated to a high Lorentz factor $\Gamma_{\rm{e j}} \sim \sigma_{0}^{3/2}$ \citep{2002A&A...387..714D}.
}

Therefore, assuming the reverse shock is mildly relativistic,  then Eq.(\ref{eq:nej1}) is reduced to}
\begin{equation}
n_{\rm{e j}} \simeq 3.7 \times 10^{7}~  E_{\rm{f,42}}^{1 / 4} \dot{m}_{\mathrm{w,-11}}^{3 / 4} \beta_{\mathrm{w,-2}}^{-3/ 4} \delta t_{-3}^{-3 / 4} a_{13}^{-3/2} \mathrm{~cm}^{-3} .
\label{eq:nej2}
\end{equation}
{How, if the  reverse shock is ultrarelativistic, this is the upper limit for $n_{\rm{e j}}$.}

{In previous studies that discussed the production of FRBs by magnetized shocks,  both forward and reverse shocks  have been involved \citep{2014MNRAS.442L...9L,2017ApJ...842...34W,2017ApJ...843L..26B,2019MNRAS.485.4091M,2020ApJ...896..142B}. In this work, we consider the synchrotron maser emission produced by both reverse shock and  forward shock. {And the equipartition of energy in the reverse and forward shock is assumed in this work for simplicity. }The theory of the synchrotron maser instability has been developed both for isotropic \citep{2002ApJ...574..861S,2019ApJ...875..126G}  and ring-like \citep{2006ApJ...652.1297L} particle distributions {in momentum space}, also see a review \cite{2021Univ....7...56L}. 
In our scenario, we consider a weakly magnetized ejecta at the deceleration radius and a stellar  wind which is itself also weakly magnetized \citep{1967ApJ...148..217W,1985A&A...152..121S,1998ApJ...505..910I,2002ApJ...576..413U,2008A&ARv..16..209P,2010ApJ...712.1157H}
\footnote{{In this regard,  the discussions in detail of the magnetization and magnetic field pattern of the ejecta and the stellar winds from the DS and their contribution to the rotation measure will be appear in a forthcoming work.}}.
%Besides, equation (\ref{eq:detagama}) shows that the upstream medium is heated only to a temperature that is insufficient to suppress the radiation efficiency of the synchrotron maser.
For the weakly magnetized plasma i.e. $\sigma \ll 1$, the characteristic frequency measured in the rest frame of the synchrotron maser may be given by \citep{2002ApJ...574..861S,2017ApJ...842...34W,2019ApJ...875..126G,2021Univ....7...56L}}
\begin{equation}
\nu_{\mathrm{c}} \approx \sigma^{-1/4} \nu_{p},
\label{eq:nuc}
\end{equation}
{where $\sigma$ is the magnetization parameter of the plasma with magnetic field of $B$ and number density of $n$,}
\begin{equation}
\sigma=\frac{B^{2}}{4 \pi n \gamma_p m_{p} c^{2}}.
\end{equation}
{The magnetization $\sigma$ is  a very uncertain and difficult parameter to know, so it can be regarded as a free parameter in principle in this work. As an example, in the case of gamma-ray bursts, after the highly magnetized jets accelerated through magnetic dissipation process, the remaining magnetization $\sigma_{\rm{ej}}$ may be small $\sim 0.1$ which can naturally account for the observed high radiative efficiency of most gamma-ray bursts \citep{2011ApJ...726...90Z}. The magnetization $\sigma_{\rm{w}}$ of the DS's wind is more difficult to determine, but perhaps we can make a very rough estimation. Suppose the DS have a surface magnetic field strength  $B_{\textborn} \sim  10^{2} ~\rm{G}$, a radius  $R_{\textborn} \sim 10 R_{\odot} $, and a mass loss rate $ \dot{m}_{\mathrm{w}} \sim 10^{-10}$ (the range of $\dot{m}_{\mathrm{w}}$ covered in this work is $10^{-9}-10^{-11}$). One gets the magnetization of the wind, at $r_{\rm{dec}}$,  $\sigma_{\rm{w}} \sim 10^{-3}$ \citep{1999isw..book.....L,2010ApJ...712.1157H}.  In addition, it can be seen from Eq.(\ref{eq:nuc}) that the result is only slightly dependent  on $\sigma$, so it is not unreasonable to directly use a rough values of $\sigma$ for orders of estimation within the range of parameters considered in this model.
}

Therefore, we  adopt  $\sigma_{\rm{ej}} = 0.1$ for the ejector at $r_{\mathrm{dec}}$ and $\sigma_{\rm{w}} = 10^{-3}$ for the DS's wind, respectively. In the context of this work, the material shocked by the reverse shock and forward shock are both baryon-dominated plasma. Therefore, the plasma frequency $\nu_{p}$ is determined  by \citep{2002ApJ...574..861S} 
\begin{equation}
\nu_{p}=\left(\frac{n_{e} e^{2}}{\pi \gamma_{e} m_{e}}+\frac{n_{p} e^{2}}{ \pi \gamma_{p} m_{p}}\right)^{1 / 2},
\end{equation}
 {where $\gamma_e$ and $\gamma_p$ are the Lorentz factor of the electrons and the protons, respectively. 
Assuming that the electrons are near equilibrium with the protons with $\gamma_{e} m_{e} \sim \gamma_{p} m_{p}$ in the downstream, we get the characteristic frequency of the synchrotron maser produced by the reverse shock and the forward shock, respectively, measured in the observed frame,}
 \begin{widetext} 
 \begin{equation}
 \nu_{\rm{pk }} \approx\left\{\begin{array}{l}
 \Gamma  \sigma_{\rm{ej}}^{-1 / 4} \left(\frac{8 n_{\rm{ej}} e^{2}}{\pi m_{p}}\right)^{1 / 2} 
 \simeq 0.88 ~\sigma_{\rm{ej,-1}}^{-1 / 4}  E_{\rm{f,42}}^{1 / 4} \dot{m}_{\mathrm{w,-11}}^{1/ 4}   \beta_{\mathrm{w,-2}}^{-1/ 4} \delta t_{-3}^{-3 / 4} a_{13}^{-1/2} ~\rm{GHz}, ~\mathrm{for ~reverse~ shock} \\ \\
 \Gamma  \sigma_{\rm{w}}^{-1 / 4}  \left(\frac{8 n_{\rm{w}} e^{2}}{\pi m_{p}}\right)^{1 / 2} 
 \simeq 0.01 ~\sigma_{\rm{w,-3}}^{-1 / 4} E_{\rm{f,42}}^{1 / 8} \dot{m}_{\mathrm{w,-11}}^{3/ 8} \beta_{\mathrm{w,-2}}^{-3/ 8} \delta t_{-3}^{-3 / 8} a_{13}^{-3/4} ~\rm{GHz}, ~\rm{for ~forward ~shock}~.
 \end{array}\right.
 \end{equation}
\end{widetext}
{ One sees that the maser emission is determined by proton, in contrast to the case for the pair plasma.}

The optical depth due to free–free absorption is estimated as
\begin{equation}
\begin{aligned}
\tau_{\mathrm{ff}} (\nu) &  \sim\frac{4 e^{6}}{3 k_{\rm{B}} m_{e} c}  (\frac{2 \pi}{3 k_{\rm{B}} m_{e}})^{1/2} \bar{g}_{\mathrm{ff}}  r_{\rm{dec}} n_{\rm{w}}^{2}(r_{\rm{dec}})  T^{-3 / 2}  \nu^{-2}\\
& \sim 10^{-8}~  \bar{g}_{\mathrm{ff}} E_{\rm{f}, 42}^{1 / 4} \dot{m}_{-11}^{7 / 4} \beta_{\rm{w},-2}^{-7/ 4} \delta t_{-3}^{1 / 4}  a_{13}^{-7/2} T_{\rm{w,4}}^{-3 / 2}  \nu_{9}^{-2},
\end{aligned}
\label{eq:detagama}
\end{equation}
where $k_{B}$ is the Boltzmann's constant, $\bar{g}_{\mathrm{ff}}$ is the mean Gaunt factor, $T_{\rm{w}}=10^4$ K is the  temperature of the wind of the DS. For $h \nu/k_{\rm{B}}T_{\rm{w}} \ll 1$,$\bar{g}_{\mathrm{ff}} \simeq (\sqrt{3}/\pi) \ln (2.2 k_{\rm{B}}T_{\rm{w}} / h \nu)$, one has $\bar{g}_{\mathrm{ff}}(\nu_{9},T_{\rm{w},4}) \simeq 7$. 
% One sees that the FRB emission is free to the free–free absorption.

The optical depth to Thomson scattering, $\tau_{\mathrm{T}} \sim \sigma_{\mathrm{T}} r_{\rm{dec}} n_{\rm{w}}(r_{\rm{dec}}) \sim 10^{-9}$, is very small.  However, due to the extremely high brightness temperature of FRBs, induced scattering process becomes important \citep{2008ApJ...682.1443L,2016ApJ...818...74L}.
Because the electrons in downstream of the shocks are ultrarelativistic, the induce Compton scattering caused by them can be negligible \citep{1978MNRAS.185..297W}. 
Therefore, the optical depth due to induced Compton scattering is mainly contributed by the DS' wind, which is given by
\begin{widetext} 
	\begin{equation}
	\tau_{\rm{I C}} \sim \frac{1}{10} ~\frac{3 \sigma_{T} c}{32 \pi m_{e}} \frac{ n_{w}\left(r_{\rm{dec}}\right) E_{\rm{FRB}}(\nu_{\rm{pk}}) }{r_{\rm{dec}}^{2} \nu_{\rm{pk}}^{3} } \simeq  \left\{\begin{array}{l}
	77~ \sigma_{\rm{ej,-1}}^{3/4} \epsilon_{-3}  E_{\rm{f},42}^{-1/4} \dot{m}_{\mathrm{w,-11}}^{3 / 4} \beta_{\mathrm{w,-2}}^{-3 / 4} \delta t_{-3}^{7/ 4} a_{13}^{-3/2} , ~\mathrm{for ~reverse~ shock} \\ \\
	10^{7}~\sigma_{\rm{w,-3}}^{3/ 4} \epsilon_{-3}  E_{\rm{f},42}^{1/8} \dot{m}_{\mathrm{w,-11}}^{3 / 8} \beta_{\mathrm{w,-2}}^{-3 / 8} \delta t_{-3}^{5/ 8} a_{13}^{-3/4} , ~\rm{for ~forward ~shock}~.
	\end{array}\right.
	\label{eq:tauIC}
	\end{equation}
\end{widetext} 
%\begin{equation}
%\begin{aligned}
%\tau_{\rm{I C}}(\nu) & \sim \frac{1}{10} ~\frac{3 \sigma_{T} c}{32 \pi m_{e}} \frac{ n_{w}\left(r_{\rm{dec}}\right) E_{\rm{FRB}} }{r_{\rm{dec}}^{2} \nu^{3} } \\ 
%& \sim  0.5~  \zeta_{-5}  E_{\rm{f},42}^{1/2} \dot{m}_{\mathrm{w,-11}}^{3 / 2} \beta_{\mathrm{w,-2}}^{-3 / 2} \delta t_{-3}^{-1/ 2} a_{13}^{-3} \nu_{9}^{-3},
%\\
%\\
%\\
%\end{aligned}
%\label{eq:tauIC}
%\end{equation}
{It can be seen that the  the emission at $\nu_{\rm{pk}}$ cannot be transmitted freely due to the induced Compton scattering. Here, $ E_{\rm{FRB}} \sim 10^{39}  \epsilon_{-3}  E_{\rm{f},42}$ erg is applied, $\epsilon$ is the efficiency of the synchrotron maser around $\nu=\nu_{\rm{pk}}$. The radiation efficiency  $\epsilon$  of synchron maser is highly uncertain. 
The PIC simulations show that the  efficiency  $\epsilon$  depends on the magnetization  and temperature for pair plasma.  \cite{2019MNRAS.485.3816P} found that  $\epsilon \sim 10^{-2}$ for an upstream magnetization $\sigma \sim 0.1$, and it decreases for $\sigma >1$. 
When the upstream plasma is non-relativistic, the  $\epsilon$ would be independent of the temperature of the plasma for $\sigma >1$ \citep{2020MNRAS.499.2884B}, but it remains unclear for $\sigma  \ll 1$.
However, for proton plasma which is the case considered in this work, the efficiency $\epsilon$  is even less known and has yet to be  investigated in detail \citep{2021Univ....7...56L}.
From the point of view of the model's consistency with observations, the efficiency should not be too low to account for the FRBs' energetics, so we assume the efficiency to be $\epsilon \sim 10^{-3}$ in this work.
}

{According to equation (\ref{eq:tauIC}), we have $\tau_{\rm{I C}}(\nu)=\tau_{\rm{I C}}(\nu_{\rm{pk}}) (\nu/\nu_{\rm{pk}})^{-(3+s)}$ if we assume the spectrum as $E_{\rm{FRB}} \propto \nu^{-s}$. 
As a result, the observed peak frequency of the radiated spectrum moves to $\nu_{\rm{m}}$ where $\tau_{\rm{I C}}(\nu_{\rm{m}})=3$,  }
\begin{widetext} 
	\begin{equation}
	\nu_{\rm{m }} \sim \left\{\begin{array}{l}
    0.88 \times 10^{1 / (3+s)} \sigma_{\rm{ej,-1}}^{-s/ 4(3+s)} \epsilon_{-3}^{1 / (3+s)} E_{\rm{f,42}}^{(2+s) / 4(3+s)} \dot{m}_{\mathrm{w,-11}}^{(6+s) / 4(3+s)} \\ \times \beta_{\mathrm{w,-2}}^{-(6+s) / 4(3+s)} \delta t_{-3}^{-(2+3 s)/ 4(3+s)} a_{13}^{-(6+s)/2(3+s)}~\mathrm{GHz}, ~\mathrm{for ~reverse~ shock} \\ \\
	 0.01 \times 10^{7/(3+s)}~\sigma_{\rm{w,-3}}^{-s / 4(3+s)} \epsilon_{-3}^{1 / (3+s)} E_{\rm{f,42}}^{(4+s) / 4(3+s)} \dot{m}_{\mathrm{w,-11}}^{3(4+s) / 8(3+s)} \\ \times \beta_{\mathrm{w,-2}}^{-3(4+s) / 8(3+s)} \delta t_{-3}^{-(4+3 s)/ 8(3+s)} a_{13}^{-3(4+s)/4(3+s)}~\mathrm{GHz}, ~\rm{for ~forward ~shock}
	\end{array}\right.
		\label{eq:vm}
	\end{equation}
\end{widetext}
{One sees that $	\nu_{\rm{m }} $ can account for the  emission of FRBs around GHz .
In addiction,{ during the reverse shock crossing phase $t < \delta t$}, this peak frequency will evolve with the shock decelerates as $\nu_{\mathrm{m}} \propto t^{-(2+3 s)/ 4(3+s)}$ for the reverse shock, and decelerates as $\nu_{\mathrm{m}} \propto t^{-(4+3 s)/ 8(3+s)}$ for the forward shock. This temporally decreasing peak frequency may explain the observed downward drifting frequency structure in the sub-pulses of some repeating FRBs \citep{2019Natur.566..235C,2019ApJ...876L..23H}. The drift rate depends on the specific value of s.}

{According to the latest observations for FRB 121102, its typical energy also $\sim 10^{37}-10^{38}$ erg \citep{2021arXiv210708205L}, and $a \sim 10^{13.5}$ cm (see section 3). {If $\dot{m}_{\mathrm{w}} =10^{-9}$, based on  Eq.(\ref{eq:Evm}), we have $\nu_{\rm{m}} \simeq 3.83~ \dot{m}_{\mathrm{w,-9}}^{15 / 36}a_{13.5}^{-5/6}~\mathrm{GHz}$ for the reverse shock, and $\nu_{\rm{m}} \simeq 1.03~ \dot{m}_{\mathrm{w,-9}}^{11 / 16} a_{13.5}^{-11/12}~\mathrm{GHz}$ for the forward shock, by adopting $s=1.5$ \citep{2019ApJ...872L..19M}.}}
This is consistent with the fact that FRB 121102 has few detection at sub-GHz band \citep{2019A&A...623A..42H,2019ApJ...882L..18J}, which may be because the DS in the case of FRB 121102 has a stronger stellar wind {of $\dot{m}_{\mathrm{w}} \sim  10^{-9}$ than the case of FRB 180916 of $ \dot{m}_{\mathrm{w}} \sim  10^{-11}$.}
%Although it may just be an instrument selection effect, which may be because the DS in the case of FRB 121102 has a stronger stellar wind of 3 than the case of  FRB 180916 of 2.

{By definition, the radiation energy  around $\nu_{\rm{pk}} $ is $E_{\rm{FRB}}(\nu_{\rm{pk}}) \equiv \epsilon  E_{\rm{f}}$, then the radiation energy  around $\nu_{\rm{m}}$ is given by $E_{\rm{FRB}}(\nu_{\rm{m}})= (\nu_{\rm{m}}/\nu_{\rm{pk}})^{-s} E_{\rm{FRB}}(\nu_{\rm{pk}})$, namely,}
\begin{widetext} 
	\begin{equation}
	E_{\rm{FRB},39}(\nu_{\rm{m }}) \approx  \left\{\begin{array}{l}
	10^{-s / (3+s)} \sigma_{\rm{ej,-1}}^{-3s/ 4(3+s)} \epsilon_{-3}^{3/ (3+s)} E_{\rm{f,42}}^{s/ 4(3+s)} \dot{m}_{\mathrm{w,-11}}^{-3s / 4(3+s)} \\ \times \beta_{\mathrm{w,-2}}^{3s / 4(3+s)} \delta t_{-3}^{-7s/ 4(3+s)} a_{13}^{3s/2(3+s)}~\mathrm{erg}, ~\mathrm{for ~reverse~ shock} \\ \\
	 10^{-7s/(3+s)}~\sigma_{\rm{w,-3}}^{-3s/ 4(3+s)} \epsilon_{-3}^{3/ (3+s)} E_{\rm{f,42}}^{-(3+s)s/ 8(3+s)} \dot{m}_{\mathrm{w,-11}}^{-3s / 8(3+s)} \\ \times \beta_{\mathrm{w,-2}}^{3s / 8(3+s)} \delta t_{-3}^{-5s/ 8(3+s)} a_{13}^{3s/4(3+s)}~\mathrm{erg}, ~\rm{for ~forward ~shock}~.
	\end{array}\right.
	\label{eq:Evm}
	\end{equation}
\end{widetext}
{Since the maser energy is mainly concentrated around $\nu_{\rm{pk}}$ in fact, the  efficiency near the observed frequency may be much lower than $10^{-3}$, which can be clearly seen from the above equation. That is, the efficiency  near the observed frequency is equal to the $\epsilon$ multiplied by a factor of less than unity, such as $5^{-s / (3+s)}$ and $10^{-7s/(3+s)}$ for reverse shock and forward shock, respectively. Then a low radiation efficiency $\sim10^{-5} ~(\ll 10^{-3})$ of the radio bursts at the observed frequency band found in the observations of the galactic burst FRB 200428 \citep{2020ApJ...899L..27M} can be understood.  
Moreover, one sees that the observed energy of the synchrotron maser  from the reverse shock is significantly greater than that from forward shock. It indicates that the observed energy may exhibit a bimodal distribution in a single repeating FRB. The  energetic bursts come from reverse shock, while the less energetic bursts come from forward shock. This may provide an explanation for the bimodal burst energy distribution in FRB 121102 found by FAST recently \citep{2021arXiv210708205L}. A more detailed analysis will be presented elsewhere.

{In this model, according to and Eq. (\ref{eq:Evm}), the bursts energy may be  adjusted mainly by the  model parameters such as $E_{\rm{f}}$, $\dot{m}_{\mathrm{w}}$ and $a$.  For different FRB sources,  there may be different model parameters, which results in different observed energies. Moreover, FRBs may be generated by both reverse and forward shock and observed by telescopes with different frequencies and thresholds, so it is very easy to generate diverse observed distributions of energy. And in order to compare directly with observations one needs to do detailed modeling, but it is beyond the scope of this work.
One thing needs special attention, Eq. (\ref{eq:vm}) and Eq. (\ref{eq:Evm}) are only valid when $\tau_{\rm{IC}} >3$. And if $\tau_{\rm{IC}} \leq  3$,  Eq. (\ref{eq:vm}) and Eq. (\ref{eq:Evm}) will  degenerate to $\nu_{\rm{m}}=\nu_{\rm{pk}}$ and $E_{\rm{FRB}, 39}=1$.
}

%\begin{equation}
%\mathcal{R}\simeq 1.2 \times 10^{7}~  E_{\rm{f,40}}^{1 / 4} \dot{m}_{\mathrm{w,-11}}^{3 / 4} \beta_{\mathrm{w,-2}}^{-3/ 4} \delta t_{-3}^{-3 / 4} a_{13}^{-3/2} \mathrm{~cm}^{-3} .
%\end{equation}
}

%{In addition to producing FRBs, the shocks also produce thermal synchrotron afterglow. Following  \cite{2019MNRAS.485.4091M}, the peak frequency of the synchrotron emission is 
%\begin{equation}
%h \nu_{\rm{s y n}} \simeq 18 ~\sigma_{-1}^{1 / 2} E_{f, 42}^{1 / 2} \delta t_{-3}^{-3 / 2}~ \rm{M e V}.
%\end{equation}
%And the peak luminosity  is  given by
%\begin{equation}
%L_{\rm{pk}} \sim 2.5 \times 10^{42}~E_{f, 40}~ \delta t_{-3}^{-1}~ \rm{erg/s},
%\end{equation}
%which is too weak to be detected at cosmological distances for existing gamma-ray detectors.
%Nevertheless, those gamma-rays can heat the upstream medium via Compton scattering.
%cite{2020MNRAS.499.2884B} demonstrated that, if the upstream medium is heated to a  high enough temperature i.e. 
%$\Delta \gamma = {k_{\rm{B}}} T_{a}/ m_{e}c^{2}  > 10^{-1.5}$, the  radiation efficiency of synchrotron maser would be strongly suppressed.  Again following the formulism in  \cite{2019MNRAS.485.4091M}, we have 
%\begin{equation}
%\begin{aligned}
%\Delta \gamma & \sim \frac{\sigma_{T} E_{f}}{32 \pi c r_{\operatorname{dec}}^{2} \sqrt{{m}_{e} h \nu_{\rm{s y n}}}} \\
%& \sim 10^{-4}~ \sigma_{-1}^{-1 / 4} E_{f, 40}^{1 / 4} \delta t_{-3}^{1 / 4} \dot{m}_{-11}^{1 / 2} \beta_{\rm{w},-2}^{-1 / 2} a_{13}^{-1},
%\end{aligned}
%\label{eq:detagama}
%\end{equation}
%which is not sufficient to suppress the synchrotron maser emission.
%} 

\section{Constraints from the periodicity and duty cycle}
Periodicities has been observed in some X-ray sources,
and mechanisms have been proposed to explain this behavior, mainly including orbital period modulation\citep{2009ApJ...706L.210S} and disk/jet precession \citep{2006MNRAS.370..399B,2010ApJ...725.2480F},
also seen in a review \cite{2017ARA&A..55..303K}.

For periodic FRBs, under the framework of our model, the periodicity might be explained  by the precession of the jets \citep{2020MNRAS.494L..64K,2021MNRAS.tmp..416K,2021arXiv210206138S}.
However, FRB 180916 shows that the signals is concentrated in a narrow active window with a duty cycle D $\simeq 0.24$, which is very different from the cases in the X-ray binaries \citep{2009ApJ...706L.210S,2010ApJ...725.2480F}.
As discussed in the previous section, FRB emission is produced when the DS fills its own Roche lobe.
Therefore, we introduce an eccentric orbit modulation mechanism to explain the narrow duty cycle. The periodicity may be explained by periodic orbital motion of the binary and the duty cycle may be explained by the eccentricity of the orbit.
It can be naturally understood by the picture:  the DS fills its Roche lobe when it is near the periastron where the flaring jets from and the FRB emission is then produced, if the orbit of the binary is eccentric.
After the DS is away from the periastron, the process of producing FRB emission stops due to significant decrease in accretion rate.

Define the mass ratio $q=M/M_{\rm{x}}$,  $M_{\rm{x}}$ is the mass of the CO, $M$ is the mass of the DS. Then the orbital period of the binary is
\begin{equation}
T=2 \pi q^{1 / 2}(1+q)^{-1 / 2} a^{3 / 2}(G M)^{-1 / 2}
\label{eq:T1}
\end{equation}
where $G$ is the gravitational constant, $a$ is the semimajor axis. The effective radius $R_{\rm{L,0}}$ of the Roche-lobe of the DS  at the periastron can be estimated as \citep{1983ApJ...268..368E}
\begin{equation}
\frac{R_{\mathrm{L,0}}}{a(1-e)}=\frac{0.49 q^{2 / 3}}{0.6 q^{2 / 3}+\ln \left(1+q^{1 / 3}\right)}\equiv \chi,
\end{equation}
where $e$ is the orbital eccentricity.
With the assumption that the DS is filled with its Roche lobe, then its average density is
\begin{equation}
\bar{\rho}=\frac{3 M}{4 \pi f_{\rm{RL}}^{3} R_{\rm{L,0}}^{3}},
\label{eq:rho}
\end{equation}
where $f_{\rm{RL}}=R/R_{\rm{L,0}} \gtrsim 1$ is the Roche lobe filling factor, $R$ is the radiu of the DS.
We expect that $f_{\rm{RL}}$ is just slightly greater than 1, .e., $f_{\rm{RL}}-1 \ll 1$. Combining equations (\ref{eq:T1})-(\ref{eq:rho}), one gets
\begin{equation}
T \simeq (3 \pi)^{1 / 2} f_{q}~(1-e)^{-3/2}(G \bar{\rho})^{-1 / 2},
\end{equation}
where $f_{q}=q^{1 / 2}(1+q)^{-1 / 2} \chi^{-3 / 2}$. Note that $f_{q} \in (1.4, 1.8)$ for $q \gtrsim 0.1$. For simplicity, we take $f_{q} = 1.5$ because the parameter range of interest in this work is $q \gtrsim 0.1$.
Thus one can simply deduce the average density of the DS as
\begin{equation}
\bar{\rho} \simeq (1-e)^{-3} \left(\frac{T}{0.21~ \rm{day}}\right)^{-2} \bar{\rho}_{\odot},
\label{eq:rho2}
\end{equation}
where $\bar{\rho}_{\odot} \simeq 1.4 ~\rm{g/cm^{3}}$ is the current average density of the Sun. It is worth noting that equation (\ref{eq:rho2}) is a generalization of equation (4.10) in \cite{2002apa..book.....F}, $\bar{\rho} \propto T^{-2}$, namely generalized from the case of circular orbit to the case of eccentric orbit. Next, we discuss the  constraint  on $e$ from the observed duty cycle D.

\begin{figure}
	\centerline{\includegraphics[angle=0,width=1.1\hsize]{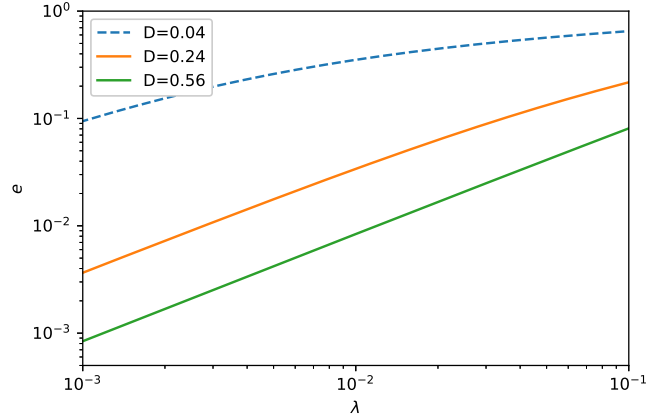}}
	\caption{The orbital eccentricity $e$ as a function of the Roche lobe overfilling parameter of the DS $\lambda$ for the duty cycle D=0.04 (FRB 180916), D=0.24 (FRB 180916) and D=0.56 (FRB 121102), respectively.}
	\label{e}
\end{figure}

Without loss of generality, taking the CO as the reference, DS moves on an elliptic orbit with respect to the CO. The distant from the CO to the DS is $r(\theta)=a(1-e^{2}) / (1+e \cos \theta)$. The DS is at periastron when $\theta=0$, where $\theta$ is the angle between the vector diameter from the CO to the DS and the polar axis in the polar coordinate system. For the DS at different positions, its Roche lobe radius is determined by $R_{\rm{L,\theta}}=\chi r(\theta)$.  Assume that when $\theta= \pm\alpha~(\alpha<\pi)$, the DS can just fill its Roche lobe, i.e., $R_{\rm{L,\alpha}}=R$. Then we have
\begin{equation}
\frac{1+e}{1+e \cos \alpha}=f_{\rm{R L}}
\end{equation}
Therefore, according to the Kepler's second law, the duty cycle of activity can be calculated as $\rm{D}=\varDelta S/S$, where $\varDelta S$ is the area swept out by $r(\theta)$ from $-\alpha$ to $\alpha$, and $S$ is the total area enclosed by the DS’s elliptic orbit. That is
\begin{equation}
D=\frac{1}{2\pi}\left(1-e^{2}\right)^{3 / 2} \int_{-\alpha}^{\alpha} \frac{d \theta}{(1+e \cos \theta)^{2}},
\end{equation}
where, by letting $\lambda=1-f_{\rm{RL}}^{-1}$,
\begin{equation}
\alpha=\left\{\begin{array}{ll}
\pi, \quad \lambda (1+e) /(2 e) \geqslant 1 \\
\\
2 \arcsin \sqrt{\lambda(1+e) /(2 e)}, \quad \lambda(1+e) /(2 e)<1
\end{array}\right.
\end{equation}
For $\lambda (1+e) /(2 e) \geqslant 1$, it means that the DS fills its Roche lobe throughout the cycle, thus D=1. Therefore, based on the observed duty cycle, we can discuss the constraint to $e$. Figure \ref{e} shows $e$ as a function of $\lambda$ for $\rm{D}=0.24$ (FRB 180916) and $\rm{D}=0.56$ (FRB 121102), respectively.
Note that $\lambda = 1-f_{\rm{RL}}^{-1}=(R-R_{\rm{L,0}})/R$ describes the degree of the Roche lobe overfilling of the DS at periastron. One naturally expects $\lambda \ll 1$, otherwise the Roche lobe overflow would be violent. By adopting $\lambda \leqslant 0.1$, we have $e \leqslant 0.22$ for FRB 180916 and $e \leqslant 0.08$ for FRB 121102, which are also showed in figure \ref{e}. It can be seen that the required orbital eccentricity is not large.

Now we can discuss what kind of DS is needed, in order to explain the periodicity of FRBs 180916 and 121102.  For FRB 180916 $T=16.35$ day and $e \leqslant 0.22$, one gets  $\bar{\rho} \sim \times 10^{-3.5} \bar{\rho}_{\odot}$ according to equation (\ref{eq:rho2}). Similarly, for FRB 121102 $T=157$ day and $e \leqslant 0.08$, one gets $\bar{\rho} \sim 10^{-6} \bar{\rho}_{\odot}$. It indicates that the DSs might be supergiants. For red supergiants, their average density can be as low as $10^{-8} \bar{\rho}_{\odot}$.
Therefore, we expect that this model can explain a period up to $T \sim 10^{3} $ day when the companion is a red supergiant.
{Note that, however,  there is some uncertainty on the phase window of FRB 180916.
It is pointed out in \cite{2020Natur.582..351C} that 50\% of the CHIME bursts are detected in a 0.6-day phase window,
with the event rate dropping rapidly towards the edges of the active phase, and the duty cycle would be D=0.04 if 0.6 day is the width of the active phase.  Moreover, \cite{2020arXiv201208348P} found that its activity window is  narrower at higher frequencies, namely,  the full-width at half-maximum of Apertif bursts is  1.1 day compared to CHIME bursts’ 2.7 day. If one adopts D=0.04, $e \leqslant 0.65$ and the density of the DS for FRB 180916 would be $\bar{\rho} \sim 10^{-2.4} \bar{\rho}_{\odot}$,  which is in line with the  massive OB stars.
}
%We note that the Roche-lobe overflow mechanism in neutron star-white dwarf binary systems had been proposed to explain the repeating FRBs\citep{2016ApJ...823L..28G,2020MNRAS.497.1543G}. However, our analysis shows that the density of white dwarf companion is too high to account for the observations.

{Moreover, it should be noted that the density obtained according to Eq.(\ref{eq:rho2}) is the average density which means that it is not necessarily the true density of DS. For example, Be stars, although they themselves are only $\sim 10R_{\odot}$, but their accretion disk radius can be as large as $\sim 100R_{\odot}$ \citep{2013A&ARv..21...69R}.  The CO can accrete material from the disk of the star, although it does not accrete material from the star itself directly \citep{2021MNRAS.tmp.1895K}. In this case, for example, for a Be star with a mass $10M_{\odot}$ and an accretion disk radius  $ \sim 10-200 R_{\odot}$, the effective average density would be $10^{-2} \bar{\rho}_{\odot}$ to $10^{-6} \bar{\rho}_{\odot}$. It can be seen that the Be companion is also consistent with the density requirements in this model. 
}

{On the other hand, according to the results in section 2, the model requires a relatively weak stellar wind environment,  specifically $\dot{M_{\rm{w}}} \sim 10^{-11}  M_{\odot}/\rm{yr}$ for FRB 180916 and $\dot{M_{\rm{w}}} \sim 10^{-9}  M_{\odot}/\rm{yr}$ for FRB 121102 may be appropriate.  It is also strongly constrained by the small DM variations observed in FRB 180916 at the low frequency bands \citep{2021ApJ...911L...3P}. Therefore, it is unlikely that the DSs are supergiants, as they tend to have much stronger wind, unless they happen to be in periods of cooling and weak wind or if the wind is highly inhomogeneous/clumpy  \citep{2008A&ARv..16..209P,2001MNRAS.325..979S}. As mentioned above, a cold OB star companion with $\dot{M_{\rm{w}}} \sim 10^{-11}  M_{\odot}/\rm{yr}$ may also reasonable  for FRB 180916. However, the DSs are more  likely to be Be stars because their polar wind may be relatively weak \citep{2006A&A...453.1059K,2008A&A...486..785K}, which could provide a unified picture for the cases of FRB 180916 and FRB 121102.  
But then again, current observations are not enough to tell us exactly what kind of stars the DSs are, and future observations are needed to provide more clues. 
In any case, this model's requirement for massive stars as DSs is consistent with the fact that FRB 121102 and FRB 180916 are {associated with the star-forming regions \citep{2017Natur.541...58C,2020Natur.577..190M,2020arXiv201103257T}.}
}
%From the previous analysis we conclude that a OB giant or supergiant which fiils its Roche lobe orbitting a stellar CO with moderate eccentricity can naturally explain the observations of the periodic FRBs, although the Roche lobe overfilling parameter $\lambda$ is unclear.
%However, neither the masses of the DS and the CO are well constrained. Nevertheless, we can take $M_{\rm{x}} = 1.4 M_{\odot}$ and $M = 30 M_{\odot}$ as an example under the assumption of $\lambda=0.1$ to
%calculate  the orbital parameters for each of the periodic FRB systems. For FRB 180916, by adopting D=0.24, we have $e \simeq 0.22$ and $\bar{\rho} \sim 10^{-3.5} \bar{\rho}_{\odot}$ ($\lambda$=0.1). Then the radius of the DS $R \simeq 45 R_{\odot}$. According to equation (\ref{eq:T1}), the  orbital semi-major axis $a \simeq 86 R_{\odot} (T/16.35{\rm{day}})^{2/3} \approx 10^{12.8}$cm, thus the apastron and periastron distances are $r_{\rm{a}}=a(1+e) \simeq 105 R_{\odot}$ and $r_{\rm{p}}=a(1-e) \simeq 67 R_{\odot}$, respectively. The Roche lobe radius of the DS at periastron $R_{\rm{L,0}}=\chi r_{\rm{p}} \simeq 42 R_{\odot}$.
%Similarly, for FRB 121102 we get $e  \simeq 0.08$, $\bar{\rho} \sim 10^{-6} \bar{\rho}_{\odot}$ and $a \simeq 386 R_{\odot} (T/157{\rm{day}})^{2/3}$.
%Therefore, $R \simeq 236 R_{\odot}$, $r_{\rm{a}} \simeq 417R_{\odot}$, $r_{\rm{p}} \simeq 355 R_{\odot}$, and  $R_{\rm{L,0}} \simeq 225 R_{\odot}$.

{ Given the presence of DS’s wind, one needs to consider its contribution to the dispersion measure (DM). 
According to Eq.(\ref{eq:nw})  and Eq.(\ref{eq:rdec}), one can roughly estimate the DM associated with the local wind, ${\rm{DM_{loc}}} \sim n_{\rm{w}}( r_{\rm{dec}}) r_{\rm{dec}} \sim 0.03~ E_{\rm{f,42}}^{1 / 4} \dot{m}_{\rm{w,-9}}^{3 / 4} \beta_{\rm{w,-2}}^{-3/ 4} \delta t_{-3}^{1 / 4} a_{13}^{-3/2}~ \rm{pc~ cm^{-3}}$ which is small enough to be negligible compared to the total DM of FRBs, even if a relatively large wind rate $\dot{m}_{\rm{w}} \sim 10^{-9}$ is adopted.  Therefore, we do not expect that  there is a obvious periodic modulation in the observed DM for FRB 180916B \citep{2020arXiv201208348P}, and of course the same is true for FRB 121102.}

{Interestingly, the DM of FRB 121102 seems to have a slow growth trend with a rare of $\sim 0.85 ~ \rm{pc~cm^{-3}~yr^{-1}}$ \citep{2021arXiv210708205L}. Based on the above analysis, it is clear that the evolution of the DS's wind is not sufficient to lead to such an outcome. The DM variation may depend on the environment in which FRB sources are located. For example, an FRB source in an expanding SNR around a nearly neutral ambient medium during the deceleration  phases or in a growing H II region can increase DM \citep{2017ApJ...847...22Y}. We're going to discuss this issue, in the context of this model,  in detail elsewhere.}

\section{Summary and discussions}
~~~~~In this work, we propose a model for understanding the highly active periodic FRBs. The system consists of a stellar CO and a DS, in which the DS fills with its own Roche lobe. {The CO accretes material from the DS and ejects relativistic magnetic blobs. The interaction between the magnetic blobs and the stellar wind of the DS produces a pair of shocks, the reverse shock traveling through the ejecta and the forward shock traveling to the wind of the DS.  We find that both of these shocks are likely to produce FRBs.  The energy of the FRBs  from the reverse shock is greater than that from forward shock. It indicates that the observed energy in a single repeating FRB may exhibit a bimodal distribution. This may provide an explanation for the bimodal burst energy distribution in FRB 121102 found by FAST recently \citep{2021arXiv210708205L}.
}

Moreover, such a Roche lobe filling accretion system can in principle sufficiently powers the highly active periodic FRBs with a long lifetime.  The orbital motion of the binary can explain the periodicity of the FRBs such as FRBs 180916 and 121102,{ if the DSs are giants/supergiants or Be stars.}
To produce a narrow duty cycle of the activity, such as FRB 180916, the orbit of the binary needs to be moderately eccentric, with the DS fills the Roche lobe only near the periastron.
It should be noted that for our model to work, it requires (1) a sufficiently large accretion rate and (2) weak stellar wind from the DS.  If not, if the accretion is too weak, the FRB energy will be too low to be observed. If the stellar wind is too strong, GHz radiation cannot pass through freely. And we realize that this is reasonable  because FRBs would have been observed in a large number of binary systems in the Milky Way if it weren't for binary systems with the right conditions to produce FRBs. 
{It is these particular low wind binary systems that produce such rare sources like FRB 180916 and FRB 121102. Therefore, if this model is correct, it gives us a great opportunity to study these special binary systems.
}

{Recent observations of FRB 180916 revealed that the bursts activity is frequency dependent, namely its activity window is both narrower and earlier at higher frequencies\citep{2020arXiv201208348P,2021ApJ...911L...3P}. The causes of this observed phenomenon may be complicated. It may be due to a combination of the absorption of FRBs by the surrounding invironment and the instrument selection effects, as show in \cite{2021arXiv210800350L}.{ It may also arise from the structured and beaming effects of the jet \citep{2021arXiv210206138S}. }But we won't discuss this issue in this work because it requires detailed modeling in the contex of this model, and we plan to study it in detail elsewhere.
}

The more recent observation found subsecond periodicity in FRB 20191221A, which may indicate that the central engine of this FRB is NS, and the period corresponds to the rotation period of NS \citep{2021arXiv210708463T}. However, the duration of this FRB is actually $\sim 3$ seconds, which is different from any known FRBs, and it has not yet been discovered whether it repeats, so it may be an entirely new class of FRBs \citep{2021arXiv210708463T}.
Nevertheless, a similar subsecond periodicity is known to exist in some X-ray binaries, {which may result from the rotation of  the accreting NSs \citep{2017ARA&A..55..303K,2021ASSL..461..143P}.
A similar periodicity is also expected in gamma-ray bursts (GRBs), although no unambiguous periodicity has been found in the GRB pulses \citep[and references therein]{2021ApJ...911...20T}.  
In this model, the rotation of the accreting NS (or the fluctuations in the accretion disc) may also modulate the accretion process and thus the generation of the jets, and whether it ultimately results in the production of periodicity in FRB pulses deserves further study. 
}

 It expects that, in the context of this model,  the energy injection by the long-term blobs ejection from the system into the surrounding environment may excite a radio nebula which can explain the persistent radio source associated with FRB 121102, {inspired by the fact that the Galactic  X-ray binaries SS433 does power such a similar radio nebula \citep{2004ASPRv..12....1F}.  }
{If this model is correct, the luminosity of the persistent radio source may be estimated as $L_{\rm{R}} \propto  \dot{E}_{\rm{FRB}}^{\gamma} $, where $\dot{E}_{\rm{FRB}} \equiv ~{E}_{\rm{FRB}}~\Re$, ${E}_{\rm{FRB}}$ is the typical energy of the FRB, $\Re$ is the repetition rate.
}
Here we adopt $\gamma=1$ for a rough estimation although the exact valu is expected to be slightly greater than 1 in the context of synchrotron radiation \citep{2017ApJ...838L...7D}.  
The typical isotropic energy $\sim 10^{37}$ erg with a repetition rate $\Re \sim 10^{-1} ~{\rm{h}}^{-1}$ for FRB 180916 \citep{2020arXiv201208348P}, and a similar isotropic energy but with a much higher repetition rate $\Re \sim 10^{2} ~{\rm{h}}^{-1}$ \citep{2021arXiv210708205L} for FRB 121102, we have $\dot{E}_{\rm{FRB ~180916}}/\dot{E}_{\rm{FRB~ 121102}} \sim 10^{-3}$. 
Therefore, we predict that the persistent radio emission associated with FRB 180916 would be $\lesssim 10^{35}$ erg/s, which is in line with observational limit \citep{2020Natur.577..190M}.

In addiction, it is expected that the accreting CO will also have persistent X-ray emission.
Assuming solar abundances, the X-ray luminosity of the accreting accretor is \citep{1973A&A....24..337S,2007MNRAS.377.1187P} 
\begin{equation}
L_{\rm{X}} \approx 1.3 \times 10^{38} \operatorname{erg} / \mathrm{s}\left\{\begin{array}{l}
\dot{m} \left(\frac{M_{\rm{x}}}{M_{\odot}}\right), \quad \dot{m} \leqslant 1 \\
(1+\ln \dot{m})\left(\frac{M_{\rm{x}}}{M_{\odot}}\right), \quad 1 \leqslant \dot{m} \leqslant 100
\end{array}\right.
\end{equation}
where $\dot{m}=\dot{M}/\dot{M}_{{\rm{Edd}}}$, the super-Eddington accretion rate $\dot{M}_{{\rm{Edd}}} \simeq 2.3 \times 10^{-8} (M_{\rm{x}}/M_{\odot}) M_{\odot} / {\rm{yr}}$. 
Based on the discussions in section 2, we have $L_{\rm{X}}  \sim 10^{38} \operatorname{erg} / \mathrm{s}$ for 
a accretion rate $\dot{M}_{a}=10^{-5} M_{\odot}/ {\rm{yr}}$, which is below the detection limit both for FRB 180916 and FRB 121102  \citep{2020ApJ...893L..42T,2020ApJ...901..165S}. 
However, we expects to detect the X-ray emission from the accretor for the sources at close distance i.e. a few tens of Mpc. 
{In addiction, we have confirmed that the prompt gamma-ray emissions radiated by the thermalized electrons  behind the (reverse and forward) shocks  associated with FRB 121102 and FRB 180916 are also expected too dim to be detected by the current gamma-ray detectors.
}

Finally, we anticipate that if those active periodic FRBs are observed at close distances in the future, multiband observations will verify or falsify our model. Also, we encourage the search for FRBs in the ultraluminous X-ray sources.

%\medskip\noindent\textit{Note added\,---\,}%%
%Just as we were about to submit this manuscript, a similar work was posted on arXiv \citep{2021arXiv210206138S}.
%We both  link periodic FRBs to the ultra-luminous X-ray  sources. In \cite{2021arXiv210206138S}, the FRBs are produced by  the interaction of the intermittent jets (high luminosity)with the quiescent jes interact with the  donor star's wind to produce FRBs. This is the key difference between our work and theirs. 
%Therefore, some physical conditions required for these two scenarios are rather different. %We are willing to discuss the difference between these two models in details  in the forthcoming work.

\section*{Acknowledgments}
This work is supported by the National Natural
Science Foundation of China (grant No. 12047550), the China Postdoctoral Science Foundation (grant No. 2020M671876)  and the  Fundamental Research Funds for the Central Universities. S.Q.Z. is supported by the China Postdoctoral Science Foundation (grant No. 2021TQ0325). ZGD was supported by the National Key Research and Development Program of China (grant No. 2017YFA0402600), the National SKA Program of China (grant No. 2020SKA0120300), and the National Natural Science Foundation of China (grant No. 11833003).

%*******************************************************************************************


\begin{thebibliography}{}
	\bibitem[Babul \& Sironi(2020)]{2020MNRAS.499.2884B} Babul, A.-N. \& Sironi, L.\ 2020, \mnras, 499, 2884. 
	\bibitem[Begelman et al.(2006)]{2006MNRAS.370..399B} Begelman, M.~C., King, A.~R., \& Pringle, J.~E.\ 2006, \mnras, 370, 399. 
	\bibitem[Beloborodov(2017)]{2017ApJ...843L..26B} Beloborodov, A.~M.\ 2017, \apjl, 843, L26. 
	\bibitem[Beloborodov(2020)]{2020ApJ...896..142B} Beloborodov, A.~M.\ 2020, \apj, 896, 142. 
	\bibitem[Bochenek et al.(2020)]{2020Natur.587...59B} Bochenek, C.~D., Ravi, V., Belov, K.~V., et al.\ 2020, \nat, 587, 59. 
	\bibitem[Caleb et al.(2019)]{2019MNRAS.484.5500C} Caleb, M., Stappers, B.~W., Rajwade, K., et al.\ 2019, \mnras, 484, 5500. 
	\bibitem[Caleb et al.(2018)]{2018NatAs...2..839C} Caleb, M., Spitler, L.~G., \& Stappers, B.~W.\ 2018, Nature Astronomy, 2, 839. 
	\bibitem[Chatterjee et al.(2017)]{2017Natur.541...58C} Chatterjee, S., Law, C.~J., Wharton, R.~S., et al.\ 2017, \nat, 541, 58. 
	\bibitem[Chime/Frb Collaboration et al.(2020)]{2020Natur.582..351C} Chime/Frb Collaboration, Amiri, M., Andersen, B.~C., et al.\ 2020, \nat, 582, 351. 
	\bibitem[CHIME/FRB Collaboration et al.(2020)]{2020Natur.587...54C} CHIME/FRB Collaboration, Andersen, B.~C., Bandura, K.~M., et al.\ 2020, \nat, 587, 54. 
	\bibitem[CHIME/FRB Collaboration et al.(2019)]{2019Natur.566..235C} CHIME/FRB Collaboration, Amiri, M., Bandura, K., et al.\ 2019, \nat, 566, 235. 
	\bibitem[Cordes \& Chatterjee(2019)]{2019ARA&A..57..417C} Cordes, J.~M. \& Chatterjee, S.\ 2019, \araa, 57, 417. 
	\bibitem[Dai \& Zhong(2020)]{2020ApJ...895L...1D} Dai, Z.~G. \& Zhong, S.~Q.\ 2020, \apjl, 895, L1. 
	\bibitem[Dai et al.(2017)]{2017ApJ...838L...7D} Dai, Z.~G., Wang, J.~S., \& Yu, Y.~W.\ 2017, \apjl, 838, L7. 
	\bibitem[Drenkhahn(2002)]{2002A&A...387..714D} Drenkhahn, G.\ 2002, \aap, 387, 714. 
	\bibitem[Eggleton(1983)]{1983ApJ...268..368E} Eggleton, P.~P.\ 1983, \apj, 268, 368. 
	\bibitem[Fabrika(2004)]{2004ASPRv..12....1F} Fabrika, S.\ 2004, \apspr, 12, 1
	\bibitem[Fender et al.(2004)]{2004MNRAS.355.1105F} Fender, R.~P., Belloni, T.~M., \& Gallo, E.\ 2004, \mnras, 355, 1105. 
	\bibitem[Fonseca et al.(2020)]{2020ApJ...891L...6F} Fonseca, E., Andersen, B.~C., Bhardwaj, M., et al.\ 2020, \apjl, 891, L6. 
	\bibitem[Foster et al.(2010)]{2010ApJ...725.2480F} Foster, D.~L., Charles, P.~A., \& Holley-Bockelmann, K.\ 2010, \apj, 725, 2480. 
	\bibitem[Frank et al.(2002)]{2002apa..book.....F} Frank, J., King, A., \& Raine, D.~J.\ 2002, Accretion Power in Astrophysics, by Juhan Frank and Andrew King and Derek Raine, pp. 398. ISBN 0521620538. Cambridge, UK: Cambridge University Press, February 2002., 398
	\bibitem[Gruzinov \& Waxman(2019)]{2019ApJ...875..126G} Gruzinov, A. \& Waxman, E.\ 2019, \apj, 875, 126. 
	\bibitem[Gu et al.(2020)]{2020MNRAS.497.1543G} Gu, W.-M., Yi, T., \& Liu, T.\ 2020, \mnras, 497, 1543. 
	%\bibitem[Gu et al.(2016)]{2016ApJ...823L..28G} Gu, W.-M., Dong, Y.-Z., Liu, T., et al.\ 2016, \apjl, 823, L28. 
	%\bibitem[Guidorzi et al.(2020)]{2020A&A...642A.160G} Guidorzi, C., Orlandini, M., Frontera, F., et al.\ 2020, \aap, 642, A160. 
	\bibitem[Harvey-Smith et al.(2010)]{2010ApJ...712.1157H} Harvey-Smith, L., Gaensler, B.~M., Kothes, R., et al.\ 2010, \apj, 712, 1157.
	\bibitem[Hessels et al.(2019)]{2019ApJ...876L..23H} Hessels, J.~W.~T., Spitler, L.~G., Seymour, A.~D., et al.\ 2019, \apjl, 876, L23. 
	\bibitem[Hjellming \& Rupen(1995)]{1995Natur.375..464H} Hjellming, R.~M. \& Rupen, M.~P.\ 1995, \nat, 375, 464. 
	\bibitem[Houben et al.(2019)]{2019A&A...623A..42H} Houben, L.~J.~M., Spitler, L.~G., ter Veen, S., et al.\ 2019, \aap, 623, A42.
	\bibitem[Ignace et al.(1998)]{1998ApJ...505..910I} Ignace, R., Cassinelli, J.~P., \& Bjorkman, J.~E.\ 1998, \apj, 505, 910. 
	\bibitem[Ioka \& Zhang(2020)]{2020ApJ...893L..26I} Ioka, K. \& Zhang, B.\ 2020, \apjl, 893, L26. 
	\bibitem[James et al.(2020)]{2020MNRAS.495.2416J} James, C.~W., Os{\l}owski, S., Flynn, C., et al.\ 2020, \mnras, 495, 2416. 
	\bibitem[Josephy et al.(2019)]{2019ApJ...882L..18J} Josephy, A., Chawla, P., Fonseca, E., et al.\ 2019, \apjl, 882, L18. 
	\bibitem[Kaaret et al.(2017)]{2017ARA&A..55..303K} Kaaret, P., Feng, H., \& Roberts, T.~P.\ 2017, \araa, 55, 303.
	\bibitem[Kanaan et al.(2008)]{2008A&A...486..785K} Kanaan, S., Meilland, A., Stee, P., et al.\ 2008, \aap, 486, 785.
	\bibitem[Karino(2021)]{2021MNRAS.tmp.1895K} Karino, S.\ 2021, \mnras, arXiv:2107.11305.
	%\bibitem[Kaspi et al.(1996)]{1996ApJ...459..717K} Kaspi, V.~M., Tauris, T.~M., \& Manchester, R.~N.\ 1996, \apj, 459, 717. 
	\bibitem[Katz(2020)]{2020MNRAS.494L..64K} Katz, J.~I.\ 2020, \mnras, 494, L64. doi:10.1093/mnrasl/slaa038
	\bibitem[Katz(2021)]{2021MNRAS.tmp..416K} Katz, J.~I.\ 2021, \mnras. doi:10.1093/mnras/stab399
	\bibitem[Kervella \& Domiciano de Souza(2006)]{2006A&A...453.1059K} Kervella, P. \& Domiciano de Souza, A.\ 2006, \aap, 453, 1059. 
	\bibitem[Lamers \& Cassinelli(1999)]{1999isw..book.....L} Lamers, H.~J.~G.~L.~M. \& Cassinelli, J.~P.\ 1999, Introduction to Stellar Winds, by Henny J. G. L. M. Lamers and Joseph P. Cassinelli, pp. 452. ISBN 0521593980. Cambridge, UK: Cambridge University Press, June 1999., 452
	\bibitem[Levin et al.(2020)]{2020ApJ...895L..30L} Levin, Y., Beloborodov, A.~M., \& Bransgrove, A.\ 2020, \apjl, 895, L30. 
	\bibitem[Li et al.(2020)]{2020arXiv200511071L} Li, C.~K., Lin, L., Xiong, S.~L., et al.\ 2020, arXiv:2005.11071
	\bibitem[Li et al.(2021)]{2021arXiv210708205L} Li, D., Wang, P., Zhu, W.~W., et al.\ 2021, arXiv:2107.08205
	\bibitem[Li et al.(2021)]{2021arXiv210800350L} Li, Q.-C., Yang, Y.-P., Wang, F.~Y., et al.\ 2021, arXiv:2108.00350
	\bibitem[Lin et al.(2020)]{2020Natur.587...63L} Lin, L., Zhang, C.~F., Wang, P., et al.\ 2020, \nat, 587, 63. 
	\bibitem[Lu et al.(2020)]{2020MNRAS.498.1397L} Lu, W., Kumar, P., \& Zhang, B.\ 2020, \mnras, 498, 1397. 
	\bibitem[Lyubarsky(2006)]{2006ApJ...652.1297L} Lyubarsky, Y.\ 2006, \apj, 652, 1297.
	\bibitem[Lyubarsky(2014)]{2014MNRAS.442L...9L} Lyubarsky, Y.\ 2014, \mnras, 442, L9. 
	\bibitem[Lyubarsky(2008)]{2008ApJ...682.1443L} Lyubarsky, Y.\ 2008, \apj, 682, 1443. 
	\bibitem[Lyubarsky \& Ostrovska(2016)]{2016ApJ...818...74L} Lyubarsky, Y. \& Ostrovska, S.\ 2016, \apj, 818, 74. 
	\bibitem[Lyutikov et al.(2020)]{2020ApJ...893L..39L} Lyutikov, M., Barkov, M.~V., \& Giannios, D.\ 2020, \apjl, 893, L39.
	\bibitem[Lyubarsky(2021)]{2021Univ....7...56L} Lyubarsky, Y.\ 2021, Universe, 7, 56.  
	\bibitem[Macquart et al.(2019)]{2019ApJ...872L..19M} Macquart, J.-P., Shannon, R.~M., Bannister, K.~W., et al.\ 2019, \apjl, 872, L19.
	\bibitem[Marcote et al.(2020)]{2020Natur.577..190M} Marcote, B., Nimmo, K., Hessels, J.~W.~T., et al.\ 2020, \nat, 577, 190.
	\bibitem[Margalit et al.(2020)]{2020ApJ...899L..27M} Margalit, B., Beniamini, P., Sridhar, N., et al.\ 2020, \apjl, 899, L27. 
	\bibitem[Mereghetti et al.(2020)]{2020ApJ...898L..29M} Mereghetti, S., Savchenko, V., Ferrigno, C., et al.\ 2020, \apjl, 898, L29. doi:10.3847/2041-8213/aba2cf
	\bibitem[Metzger et al.(2019)]{2019MNRAS.485.4091M} Metzger, B.~D., Margalit, B., \& Sironi, L.\ 2019, \mnras, 485, 4091. 
	\bibitem[Mottez et al.(2020)]{2020A&A...644A.145M} Mottez, F., Zarka, P., \& Voisin, G.\ 2020, \aap, 644, A145. 
	\bibitem[Palaniswamy et al.(2018)]{2018ApJ...854L..12P} Palaniswamy, D., Li, Y., \& Zhang, B.\ 2018, \apjl, 854, L12. 
	\bibitem[Patruno \& Watts(2021)]{2021ASSL..461..143P} Patruno, A. \& Watts, A.~L.\ 2021, Astrophysics and Space Science Library, 461, 143.
	\bibitem[Pastor-Marazuela et al.(2020)]{2020arXiv201208348P} Pastor-Marazuela, I., Connor, L., van Leeuwen, J., et al.\ 2020, arXiv:2012.08348
	\bibitem[Petroff et al.(2019)]{2019A&ARv..27....4P} Petroff, E., Hessels, J.~W.~T., \& Lorimer, D.~R.\ 2019, \aapr, 27, 4. 
	\bibitem[Pleunis et al.(2021)]{2021ApJ...911L...3P} Pleunis, Z., Michilli, D., Bassa, C.~G., et al.\ 2021, \apjl, 911, L3.
	\bibitem[Plotnikov \& Sironi(2019)]{2019MNRAS.485.3816P} Plotnikov, I. \& Sironi, L.\ 2019, \mnras, 485, 3816. 
	\bibitem[Poutanen et al.(2007)]{2007MNRAS.377.1187P} Poutanen, J., Lipunova, G., Fabrika, S., et al.\ 2007, \mnras, 377, 1187. 
	\bibitem[Puls et al.(2008)]{2008A&ARv..16..209P} Puls, J., Vink, J.~S., \& Najarro, F.\ 2008, \aapr, 16, 209. 
	\bibitem[Rajwade et al.(2020)]{2020MNRAS.495.3551R} Rajwade, K.~M., Mickaliger, M.~B., Stappers, B.~W., et al.\ 2020, \mnras, 495, 3551. doi:10.1093/mnras/staa1237
	\bibitem[Ridnaia et al.(2020)]{2020arXiv200511178R} Ridnaia, A., Svinkin, D., Frederiks, D., et al.\ 2020, arXiv:2005.11178
	\bibitem[Rivinius et al.(2013)]{2013A&ARv..21...69R} Rivinius, T., Carciofi, A.~C., \& Martayan, C.\ 2013, \aapr, 21, 69
	\bibitem[Sagiv \& Waxman(2002)]{2002ApJ...574..861S} Sagiv, A. \& Waxman, E.\ 2002, \apj, 574, 861. 
	\bibitem[Sakurai(1985)]{1985A&A...152..121S} Sakurai, T.\ 1985, \aap, 152, 121
	\bibitem[Sari \& Piran(1995)]{1995ApJ...455L.143S} Sari, R. \& Piran, T.\ 1995, \apjl, 455, L143. doi:10.1086/309835
	\bibitem[Scholz et al.(2020)]{2020ApJ...901..165S} Scholz, P., Cook, A., Cruces, M., et al.\ 2020, \apj, 901, 165. 
	\bibitem[Shakura \& Sunyaev(1973)]{1973A&A....24..337S} Shakura, N.~I. \& Sunyaev, R.~A.\ 1973, \aap, 500, 33
	\bibitem[Sironi \& Spitkovsky(2011)]{2011ApJ...726...75S} Sironi, L. \& Spitkovsky, A.\ 2011, \apj, 726, 75
	\bibitem[Spitler et al.(2016)]{2016Natur.531..202S} Spitler, L.~G., Scholz, P., Hessels, J.~W.~T., et al.\ 2016, \nat, 531, 202. 
	\bibitem[Sridhar et al.(2021)]{2021arXiv210206138S} Sridhar, N., Metzger, B.~D., Beniamini, P., et al.\ 2021, arXiv:2102.06138
	\bibitem[Stairs et al.(2001)]{2001MNRAS.325..979S} Stairs, I.~H., Manchester, R.~N., Lyne, A.~G., et al.\ 2001, \mnras, 325, 979.
	\bibitem[Strohmayer(2009)]{2009ApJ...706L.210S} Strohmayer, T.~E.\ 2009, \apjl, 706, L210. 
	\bibitem[Tarnopolski \& Marchenko(2021)]{2021ApJ...911...20T} Tarnopolski, M. \& Marchenko, V.\ 2021, \apj, 911, 20. 
	\bibitem[Tavani et al.(2020)]{2020arXiv200512164T} Tavani, M., Casentini, C., Ursi, A., et al.\ 2020, arXiv:2005.12164
	\bibitem[Tavani et al.(2020)]{2020ApJ...893L..42T} Tavani, M., Verrecchia, F., Casentini, C., et al.\ 2020, \apjl, 893, L42. 
	\bibitem[Tendulkar et al.(2020)]{2020arXiv201103257T} Tendulkar, S.~P., Gil de Paz, A., Kirichenko, A.~Y., et al.\ 2020, arXiv:2011.03257
	\bibitem[The CHIME/FRB Collaboration et al.(2021)]{2021arXiv210708463T} The CHIME/FRB Collaboration, Andersen, B.~C., Bandura, K., et al.\ 2021, arXiv:2107.08463
	\bibitem[Tong et al.(2020)]{2020RAA....20..142T} Tong, H., Wang, W., \& Wang, H.-G.\ 2020, Research in Astronomy and Astrophysics, 20, 142. doi:10.1088/1674-4527/20/9/142
	\bibitem[ud-Doula \& Owocki(2002)]{2002ApJ...576..413U} ud-Doula, A. \& Owocki, S.~P.\ 2002, \apj, 576, 413. 
	\bibitem[Waxman(2017)]{2017ApJ...842...34W} Waxman, E.\ 2017, \apj, 842, 34. 
	\bibitem[Weber \& Davis(1967)]{1967ApJ...148..217W} Weber, E.~J. \& Davis, L.\ 1967, \apj, 148, 217. 
	\bibitem[Wilson \& Rees(1978)]{1978MNRAS.185..297W} Wilson, D.~B. \& Rees, M.~J.\ 1978, \mnras, 185, 297.
	\bibitem[Wiktorowicz et al.(2015)]{2015ApJ...810...20W} Wiktorowicz, G., Sobolewska, M., S{{a}}dowski, A., et al.\ 2015, \apj, 810, 20. 
	\bibitem[Xiao et al.(2021)]{2021SCPMA..6449501X} Xiao, D., Wang, F., \& Dai, Z.\ 2021, Science China Physics, Mechanics, and Astronomy, 64, 249501.
	\bibitem[Yang \& Zou(2020)]{2020ApJ...893L..31Y} Yang, H. \& Zou, Y.-C.\ 2020, \apjl, 893, L31. 
	\bibitem[Yang \& Zhang(2017)]{2017ApJ...847...22Y} Yang, Y.-P. \& Zhang, B.\ 2017, \apj, 847, 22.
	\bibitem[Yuan et al.(2009)]{2009MNRAS.395.2183Y} Yuan, F., Lin, J., Wu, K., et al.\ 2009, \mnras, 395, 2183. 
	\bibitem[Yuan \& Zhang(2012)]{2012ApJ...757...56Y} Yuan, F. \& Zhang, B.\ 2012, \apj, 757, 56. 
	\bibitem[Zanazzi \& Lai(2020)]{2020ApJ...892L..15Z} Zanazzi, J.~J. \& Lai, D.\ 2020, \apjl, 892, L15. 
	\bibitem[Zhang(2020)]{2020ApJ...890L..24Z} Zhang, B.\ 2020, \apjl, 890, L24. 
	\bibitem[Zhang(2020)]{2020Natur.587...45Z} Zhang, B.\ 2020, \nat, 587, 45. 
	\bibitem[Zhang \& Yan(2011)]{2011ApJ...726...90Z} Zhang, B. \& Yan, H.\ 2011, \apj, 726, 90.
	\bibitem[Zhang \& Yu(2015)]{2015MNRAS.451.1740Z} Zhang, H. \& Yu, W.\ 2015, \mnras, 451, 1740. 
	\bibitem[Zhao et al.(2020)]{2020MNRAS.499.1561Z} Zhao, T.-L., Yuan, Y.-F., \& Kumar, R.\ 2020, \mnras, 499, 1561. 
	\bibitem[Zhong et al.(2020)]{2020ApJ...898L...5Z} Zhong, S.-Q., Dai, Z.-G., Zhang, H.-M., et al.\ 2020, \apjl, 898, L5. 
	
	
	
	
	
\end{thebibliography}
\end{document}